\documentclass[a4paper,10pt]{article}
\usepackage{graphicx}
\usepackage{epstopdf}
\usepackage{amsfonts,amsmath,amssymb}
\usepackage{hyperref}

\usepackage{epsfig,multicol}

\newcommand\fverb{\setbox\pippobox=\hbox\bgroup\verb}
\newcommand\fverbit{\egroup\item[\fbox{\unhbox\pippobox}]}

\newbox\pippobox

\begin{document}
\title{\bf  Dark Matter From $f(R,T)$ Gravity}
\author{Raziyeh Zaregonbadi\thanks{Electronic address: r$\_$zare@sbu.ac.ir}\, ,\,
        Mehrdad Farhoudi\thanks{Electronic address:
        m-farhoudi@sbu.ac.ir}\, and\, Nematollah Riazi\thanks{Electronic
        address: n$\_$riazi@sbu.ac.ir}\, \,
\\
\small Department of Physics, Shahid Beheshti University, G.C., Evin, Tehran 19839, Iran}
\date{\small October 25, 2016}
\maketitle
\begin{abstract}
\noindent
 We consider $f\left( {R,T} \right)$ modified theory of
gravity, in which the gravitational Lagrangian is given by an
arbitrary function of the Ricci scalar and the trace of the
energy--momentum tensor of the matter, in order to investigate the
dark--matter effects on the galaxy scale. We obtain the metric
components for a spherically symmetric and static spacetime in the
vicinity of general relativity solutions. However, we concentrate
on a specific model of the theory where the matter is minimally
coupled to the geometry, and derive the metric components in the
galactic halo. Then, we fix the components by the rotational
velocities of the galaxies for the model, and show that the mass
corresponding to the interaction term (which appears in the
Einstein modified field equation) leads to a flat rotation curve
in the halo of galaxies. In addition, for the proposed model, the
light--deflection angle has been derived and drawn using some
observed data.
\end{abstract}

\bigskip
\section{Introduction}\label{sec1}
\indent

The issue of dark matter has been a long--outstanding problem in
cosmology and galactic astronomy. The observational aspects, such
as the behavior of the galactic rotation curves and the mass
discrepancy in clusters of galaxies, have necessarily led to the
consideration of the existence of dark matter~\cite{1}. According
to the Newtonian gravity, the galactic rotation curves require the
velocity of a star or an interstellar cloud, rotating in the disk,
to linearly increase within the bulge and to drop off as the
square root of $ 1/r $ in the outer parts. In contrast, the
observed rotation curves of spiral galaxies show that even though
the rotational velocities increase from the center of the galaxy,
they then attain approximately constant values in the outer range
of the baryonic matter's disk (up to several luminous
radii\rlap,\footnote{However, beyond the range of nearly flat
rotational velocity, there is some observational evidence showing
the decay of the galactic rotation curves, see, e.g.,
Ref.~\cite{Huang2016}.}\
 see, e.g., Ref.~\cite{2}). Such evidence
indicates the possible existence of some new invisible matter
distributed inside and (mostly) around galaxies, which is known as
dark matter~\cite{2,3}.

In this respect, it is well known that the mass of a cluster can
be estimated in two ways. The total baryonic mass, $ M_b $, is
estimated by considering the total sum of all observable mass
members. However, by taking into account the dynamical motions of
galaxies, the virial theorem provides an estimated mass, $ M $,
for a cluster. The comparison between the total baryonic mass
(obtained from the observed data) with the virial mass of a
cluster shows that their ratio is $ M/M_b\approx 20-30 $. This
fact is usually explained by postulating the existence of dark
matter~\cite{1,4}. In addition, the curvature of the spacetime,
near any gravitating mass (including the dark matter), deflects
passing rays of light, and distorts the images of the background
galaxies. Indeed, gravity acts as a lens to bend the light from a
more distant source (such as a quasar) around a massive object
(such as a cluster of galaxies) lying between the source and the
observer, in accordance with general relativity
(GR)~\cite{Einstein}. Such an effect was first observed in 1919,
during a solar eclipse in front of the Hyades star cluster, whose
stars appeared to move as they passed behind the sun~\cite{Dyson}.
However, this concept was improved in $1937$, when Zwicky
suggested that the ultimate measurement of the cluster masses
would emerge from gravitational lensing~\cite{Zwicky}; this has
indeed become the most successful probe of the dark sector.
Nowadays, the measurements of such effects provide constraints on
the mean density of dark matter~\cite{Hoekstra}.

There are several candidates for dark matter. One possible
categorization tells us that dark matter can be either baryonic or
non--baryonic. The baryonic sector is made of baryons (protons and
neutrons) that make up stars, interstellar matter and planets. The
main baryonic candidates are the massive astronomical compact halo
objects (MACHOs) that include brown dwarf stars and black holes.
The non--baryonic candidates are basically elementary particles
that have non--standard properties. Among the non--baryonic
candidates, one can point to the axions, which are considered as a
solution to the strong--CP problem. However, the largest class is
that of the weakly interacting massive particle (WIMP), which can
be various types of unknown particles~\cite{5}. The most popular
of these WIMPs is the neutralino from supersymmetry. The WIMP
interaction cross--section with the normal baryonic matter is
extremely small, but non--zero; thus, their direct detection is a
weak possibility. Heavy neutrinos may also be considered as
possible candidates for dark matter~\cite{5}.

Moreover, accelerator and reactor experiments do~not yet support
the particle scenarios in which dark matter emerges. To deal with
the question of dark matter, a great number of alternative efforts
have been concentrated on various modifications to the Einstein
field equations (see, e.g., Refs.~\cite{6}--\cite{Harko2007}). For
instance, theories of $f(R)$--modified gravity (as the simplest
family of the higher--order gravities), which are based on
replacing the scalar curvature $ R $ in the Einstein--Hilbert
action with an arbitrary differentiable function of it, have had
some successes in explaining the accelerated expansion of the
universe~\cite{CNOT}--\cite{Atazadeh2008}, and have accounted for
the dark--matter--like effects~\cite{18}--\cite{Sefidgar}.

In this work, we propose to explain the effect related to dark
matter by another type of modified gravity theory, namely,
$f(R,T)$ gravity. The theory of $ f(R,T)$ gravity generalizes
theories of gravity by {\it a priori} incorporation of the trace
of dustlike matter in addition to the Ricci scalar into the
Lagrangian (see, e.g.,
Refs.~\cite{Harko-2008}--\cite{shabani16b}). That is, the
Lagrangian of $ f(R,T) $ gravity model depends on a source term,
which itself represents the variation of the energy--momentum
tensor with respect to the metric. In other words, the appearance
of the matter with an unusual coupling with the geometry may have
a relation with the geometrical curvature induction of the matter
in addition to the pure geometry in spacetime. Hence, we
investigate whether the interaction between the matter and
geometry can explain the observational data in the galaxy halo
instead of considering an additional mysterious mass as dark
matter.

The work is organized as follows. The field equations in $f(R,T)$
gravity, in particular when the matter is minimally coupled to the
curvature in a specific form, are presented in Sec.~$2$, where we
also specify the metric components for this type of theory in a
spherical galactic halo. In Sec.~$3$, we fix the metric components
while using the tangential velocity of a galaxy. Then, we study
the propagation of the light in this type of theory for a typical
galaxy in Sec.~$4$. Finally, in Sec.~$5$ we present the
conclusions. Through the work, we use the sign convention $( { -
,+ , + , + })$
 and geometrical units with $c=1$.
\section{Modified Field Equations of $f(R,T)$ Gravity}\label{sec 2}
\indent

In this section, we derive the field equations and some
corresponding dynamical parameters of $f(R,T)$--modified gravity
in four--dimensional spacetime. The action is simply written in
the form
\begin{equation}\label{eq1}
S = \int {{d^4}x\sqrt { - g} \left[ {\frac{1}{{2\kappa
}}f\left({R,T} \right) + {L_m}} \right]},
\end{equation}
where $ \kappa  \equiv 8\pi G $, $g$ is the determinant of the
metric and $ {L_m} $  is the matter Lagrangian density. The
energy--momentum tensor is defined as
\begin{equation}\label{eq2}
T_{\mu \nu }^{[m]} =  - \frac{2}{{\sqrt { - g} }}\frac{{\delta(
{\sqrt { - g}\, {L_m}})}}{{\delta {g^{\mu \nu }}}},
\end{equation}
where the lowercase Greek indices run from zero to three and the
index $m$ stands for the baryonic matter. The variation of the
action with respect to the metric tensor gives the field equations
\begin{equation}\label{eq3}
{f_R}{R_{\mu \nu }} - \frac{f}{2}{g_{\mu \nu }} + \left({g_{\mu
\nu }{\Box}}- {\nabla _\mu }{\nabla _\nu }\right){f_R} = \left(
{\kappa + {f_T}} \right)T_{\mu \nu }^{[m]},
\end{equation}
where $ \Box \equiv {\nabla _\mu }{\nabla ^\mu } $ and we have
defined the following functions for the derivative of the function
$ f(R,T) $ with respect to its arguments, i.e.,
\begin{equation}\label{eq4}
{f_R} \equiv \frac{{\partial f(R,T)}}{{\partial R}}
\qquad\qquad\quad {\rm and} \qquad\qquad\quad {f_T} \equiv
\frac{{\partial f(R,T)}}{{\partial T}}.
\end{equation}

It is usually more instructive to write the field equations in the
form of the Einstein equations with an effective energy--momentum
tensor as
\begin{equation}\label{eq6}
{G_{\mu \nu }} = \frac{\kappa }{{{f_R}}}\left( T_{\mu \nu }^{[m]}+
T_{\mu \nu }^{{\mathop{[\rm int]}} } \right) =\frac{\kappa }{{{f_R}}}T_{\mu \nu }^{[\rm eff]},
\end{equation}
where $T_{\mu \nu}^{[\rm eff]} \equiv T_{\mu \nu }^{[m]} + T_{\mu
\nu}^{[{\mathop{\rm int}} ]}$. The interaction energy--momentum
tensor has been defined as
\begin{equation}\label{eq7}
T_{\mu \nu }^{{\mathop{[\rm int]}} } \equiv \frac{1}{\kappa
}\left[{{f_T}T_{\mu \nu }^{[m]} + \frac{1}{2}(f - R{f_R}){g_{\mu
\nu }} + \left( {{\nabla _\mu }{\nabla _\nu } - {g_{\mu \nu
}{\Box}}} \right){f_R}} \right],
\end{equation}
which may be interpreted as a fluid composed of the interaction
between the matter and the curvature terms.

In the following subsections, we first derive the field equations
for a spherically symmetric system. Then, we study a particular
minimal coupling model in $f(R,T)$ gravity to investigate the
dark--matter effect on the galactic scale, wherein we also obtain
the metric components for this model.

\subsection{Field Equations For Spherically Symmetric System}
\indent

Let us now consider an isolated system that is described by a
static and spherically symmetric metric
\begin{equation}\label{metric}
d{s^2} =  - {e^{a\left( r \right)}}d{t^2} + {e^{b\left( r \right)}}d{r^2} + {r^2}d{\theta ^2} + {r^2}{\sin ^2}\theta d{\phi ^2}.
\end{equation}
Thus, the energy--momentum tensor of the matter can be described
by an effective density, $ {\rho ^{[\rm eff]}} $, and an effective
anisotropic pressure with radial, $ {p_r^{[\rm eff]}} $, and
tangential, ${p_ \bot ^{[\rm eff]}} $,
components~\cite{Jackson1970}. Henceforth, the field equations
become
\begin{equation}\label{field eq1}
G_t^t =  - \frac{1}{{{r^2}}} + \frac{1}{{{r^2}{e^b}}} - \frac{{b'}}{{r{e^b}}} =  -\frac{\kappa }{{{f_R}}}
\left( {{\rho ^{[m]}} + {\rho ^{\left[ {{\mathop{\rm int}} } \right]}}} \right),
\end{equation}
\begin{equation}\label{field eq2}
G_r^r =  - \frac{1}{{{r^2}}} + \frac{1}{{{r^2}{e^b}}} + \frac{{a'}}{{r{e^b}}} =
\frac{\kappa }{{{f_R}}}p_r^{\left[ {{\mathop{\rm int}} } \right]}
\end{equation}
and
\begin{equation}\label{field eq3}
G_\theta ^\theta  = G_\phi ^\phi  = \frac{1}{{4{e^b}}}\left( {2a'' + {{a'}^2} - a'b'} \right) +
\frac{{a' - b'}}{{2r{e^b}}} = \frac{\kappa }{{{f_R}}}p_ \bot ^{\left[ {{\mathop{\rm int}} } \right]}.
\end{equation}
From definition (\ref{eq7}), we get the energy density and the  radial pressure
for the interaction term as
\begin{equation}\label{density}
{\rho ^{\left[ {{\mathop{\rm int}} } \right]}} =  - T_t^{t\left[ {{\mathop{\rm int}} } \right]} =
\frac{1}{\kappa }\left[ {{f_T}{\rho ^{\left[ m \right]}} + \frac{1}{2}\left( {R{f_R} - f} \right) -
\frac{{a'{f'_R}}}{{2{e^b}}} +\Box {f_R}} \right],
\end{equation}
\begin{equation}\label{press}
p_r^{[ {{\mathop{\rm int}} }]} = T_r^{r\left[ {{\mathop{\rm int}} } \right]}=
\frac{1}{\kappa }\left[ {\frac{1}{2}\left( {f - R{f_R}} \right) - \frac{1}{{{e^b}}}
\left( {\frac{{b'}}{2}{f'_R} - {f''_R}} \right) -\Box {f_R}} \right],
\end{equation}
\begin{equation}\label{pressteta}
p_{\bot }^{[ {{\mathop{\rm int}} }]} =T_\theta ^{\theta \left[ {\operatorname{int} } \right]} =
T_\phi ^{\phi \left[ {\operatorname{int} } \right]} =
 \frac{1}{\kappa }\left[ {\frac{1}{2}\left( {f - R{f_R}} \right) + \frac{{{f'_R}}}{{r{e^b}}} -\Box {f_R}} \right],
\end{equation}
where prime is derivative with respect to the $ r $--coordinate.
In these relations, $ \Box {f_{R}} $ is
\begin{equation}
\Box {f_{R}}=\frac{1}{{{e^b}}}\left[ {\left( {\frac{{a' - b'}}{2}} \right){f'_R} + \frac{2}{r}{f'_R} + {f''_R}} \right] .
\end{equation}

To obtain the components of the metric for this type of modified
gravity, we first derive a useful relation from  metric
({\ref{metric}), namely
\begin{equation}\label{13}
 \frac{R_{tt}}{e^{a}}+ \frac{R_{rr}}{e^{b}}=\frac{{a' + b'}}{{r{e^b}}},
\end{equation}
as well as another one, from the field equations (\ref{eq3}), as
\begin{equation}\label{14}
  \frac{R_{tt}}{e^{a}}+ \frac{R_{rr}}{e^{b}}=\frac{1}{{{f_R}}}\left[ {\left( {\kappa  +
  {f_T}} \right){\rho ^{\left[ m \right]}} - \frac{{{f'_R}}}{{2{e^b}}}\left( {a' + b'} \right) +
  \frac{{{f''_R}}}{{{e^b}}}} \right].
\end{equation}
Replacing relations (\ref{density}) and (\ref{press}) into the
right--hand side of relation (\ref{14}), and then inserting
relation (\ref{14}) into (\ref{13}) gives
\begin{equation}\label{15}
\frac{{a' + b'}}{{r{e^b}}} = \frac{\kappa }{{{f_R}}}\left( {\rho
^{\left[ m \right]}} + {\rho ^{\left[ {{\mathop{\rm int}} }
\right]}} + {p_{r}^{\left[ {{\mathop{\rm int}} } \right]}}
\right).
\end{equation}
On the other hand, our purpose is to obtain solutions that differ
from the classical GR only slightly. In this respect, if the
combination $ a' + b' $ is a well--behaved differential
expression, it should have a solution of the form $
{e^{a(r)}}{e^{b(r)}} = A(r) $; however, in order to remain in the
vicinity of GR, the function $ A(r) $ would slightly be different
from $1$. For instance, one can consider~\cite{sobouti}
\begin{equation}\label{assum}
 A\left( r \right) = {\left( {r/s} \right)^\beta } \qquad \Rightarrow \qquad a' + b' = \frac{\beta }{r},
\end{equation}
where $ \beta $ is a dimensionless parameter and $ s $ is the
length scale of the system; however, if $\beta \ll 1$, one gets
$A(r) \approx 1 + \beta \ln \left( {r/s} \right)$ and thus, the
desired proposal will be fulfilled. By inserting relation
(\ref{assum}) into (\ref{15}), we can specify the components of
the metric as
\begin{equation}\label{eB}
{e^{b\left( r \right)}} = \frac{{\beta {f_R}}}{{\kappa {r^2}\left( {{\rho ^{\left[ m \right]}} +
{\rho ^{\left[ {\operatorname{int} } \right]}} + p_r^{\left[ {\operatorname{int} } \right]}} \right)}}
\end{equation}
and
\begin{equation}\label{eA}
{e^{a\left( r \right)}} = {\left( {\frac{r}{s}} \right)^\beta }{e^{ - b\left( r \right)}}.
\end{equation}

\subsection{Minimal Coupling Model}
\indent

Among different types of modified $f(R,T)$ gravity, inspired from
our previous work~\cite{Zare}, we consider a model where the
matter, in a simple manner, is minimally coupled to the geometry
as
\begin{equation}\label{16}
f(R,T) = R - \alpha(- T)^{n},
\end{equation}
where $ T=-\rho^{[m]} $, and (positive) $\alpha $ and $n$ are
constants, wherein $n$ is a power which determines the strength of
the effect of the matter. Furthermore, it is worth mentioning that
beyond the radius of the galactic halo, if there is a true vacuum
status (i.e., $T_{\mu\nu}^{[m]}=0$), then the model will obviously
reduce to GR. For this model, we have
\begin{equation}\label{17}
{f_R} = 1 \qquad\qquad\quad {\rm  and} \qquad\qquad\quad
{f_T} =\alpha n{\left({\rho^{[m]}}\right)^{n - 1}}.
\end{equation}
Hence, as the interaction energy density, relation
(\ref{density}), gives
\begin{equation}\label{rho}
\kappa {\rho ^{\left[ {\operatorname{int} } \right]}} = \alpha
\left( {n + \frac{1}{2}} \right){\left( {\rho^{[m]}} \right)^n},
\end{equation}
a plausible requirement $ \rho^{[\rm int]}\geqslant 0$, with
positive $ \alpha $, implies $ n\geqslant -1/2$. However, in order
to have a solution for the vacuum case, the power of the matter
density should be equal or more than zero, i.e. $ n\geqslant0 $,
where $ n=0 $ corresponds to GR theory. Also, from relations
(\ref{density}) and (\ref{press}), we get
\begin{equation}\label{ro+p}
\kappa \left( {{\rho ^{\left[ {\operatorname{int} } \right]}} + p_r^{\left[ {\operatorname{int} }
\right]}} \right) = \alpha n{\left( {\rho^{[m]}} \right)^n}.
\end{equation}

In the galactic halo, where the density of the baryonic matter is
very low, we assume that the mass assigned to the density of the
interaction between the matter and geometry can explain the
observations with no need to introduce a mysterious dark matter.
In fact, with $0<n<1$, relation (\ref{rho}) still yields $
\rho^{[m]}\ll\rho^{[\rm int]}$ in this region. Indeed, where there
is very little matter, but as far as $ T^{[m]} $ is~not exactly
zero (although, $T^{[m]}\rightarrow 0$), the model can, in
principle, maintain the flatness of the rotation curves.

Thus, by assuming $ \rho^{[m]}\ll\rho^{[\rm int]}$ in the galactic
halo and inserting relation (\ref{ro+p}) into relation (\ref{eB}),
we can rewrite the component of the metric as
\begin{equation}\label{eb}
{e^{b\left( r \right)}} = \frac{\beta }{{ \alpha n{r^2}{{\left( {{\rho ^{\left[ m \right]}}} \right)}^n}}}.
\end{equation}
In addition, although the baryonic matter in the form of stars,
interstellar gas and dust in spiral galaxies is mostly
concentrated in the flattened disk, it is instructive to suppose
that, in the galactic halo, the density distribution of the
baryonic matter is approximated by a spherically symmetric model
and rapidly decreases with radius according to the power--law
profile\footnote{Note that, we are~not building a fully realistic
model, hence, we have restricted ourselves to a spherically
symmetric approximation, as many other authors, see, e.g.,
Refs.~\cite{Keeton97,Bohmer08}.}
\begin{equation}
\rho^{[m]}=\rho_{0}\, {r^{-\gamma}},
\end{equation}
where\footnote{Note that, according to the Newtonian gravity, one
has to have a constant rotational velocity in the halo of the
galaxy, i.e., $M \propto r \quad\Rightarrow\quad \rho^{[m]}
\propto {r^{ - 2}} $ .}
 $ \gamma >2 $ and $ \rho_{0} $ is a constant that, without loss of generality,
we set equal to $ 1 $. Also, we take the length scale of the
system to be equal to the radius of the baryonic matter dominated,
i.e. $ s=r_B $. Hence, from relation (\ref{eb}), we get
\begin{equation}\label{ebx}
{e^{b\left( r \right)}} = \frac{\beta }{\alpha n}{r^{\gamma n -
2}}
\end{equation}
and, consequently, from relation (\ref{eA}),
\begin{equation}\label{eax}
{e^{a\left( r \right)}} = \frac{{{ \alpha n}}}{{\beta {r^{\beta}
_{B} }}}{r^{\beta  + 2 - \gamma n}}.
\end{equation}
The interaction density, relation (\ref{rho}), reads
\begin{equation}\label{roint}
\kappa {\rho^{[\rm int]}}=\alpha \left( {n + \frac{1}{2}}
\right){r^{ - \gamma n}}.
\end{equation}

In the next section, we attempt to fix the metric components
(\ref{ebx}) and (\ref{eax}) for a galactic model with the flat
rotation curve.

\section{Galactic Rotation Curves}
\indent

The observational data shows that the rotational velocity
increases linearly within the bulge and approaches a constant
value of about $ 200-500$ km/s, as one moves away from the core up
to several luminous radii~\cite{1,2}. In this section, we consider
a test particle that moves in a timelike geodesic orbit in a
static and spherically symmetric system in the plane (without loss
of generality) $ \theta = \pi /2 $. Hence, the corresponding
geodesic equation for the $ r $--coordinate is
\begin{equation}\label{geodesic}
\frac{{{d^2}r}}{{d{\tau ^2}}} + \frac{{a'{e^{a\left( r \right)}}}}{{2{e^{b\left( r \right)}}}}{
\left( {\frac{{dt}}{{d\tau }}} \right)^2} + \frac{{b'}}{2}{\left( {\frac{{dr}}{{d\tau }}}
\right)^2} - \frac{r}{{{e^b}}}{\left( {\frac{{d\phi }}{{d\tau }}} \right)^2} = 0,
\end{equation}
where $ \tau $ is the affine parameter along the geodesic. When a
test particle moves along its geodesic in a gravitational field in
a spherically symmetric spacetime, the momenta $ P_{0} $ and $
P_{3} $ are conserved~\cite{Weinberg,Straumann}, namely,
\begin{equation}\label{const}
E = {e^{a\left( r \right)}}\left( {\frac{{dt}}{{d\tau }}} \right)
={\rm const}. \qquad\qquad {\rm and} \qquad\qquad J= {r^2}\left(
{\frac{{d\phi }}{{d\tau }}} \right) = {\rm const}.,
\end{equation}
where $ E $ is the energy and $ J $ is the $ \phi $--coordinate of
the angular momentum of the test particle.

Now, let us study the motion associated with the stable circular
orbits, i.e. $ dr/d\tau  = 0$. In this case, we can rewrite Eq.
(\ref{geodesic}) as
\begin{equation}\label{geodesic2}
\frac{{a'{E^2}}}{{2{e^{a\left( r \right)}}}} = \frac{{{J^2}}}{{{r^3}}}.
\end{equation}
In the weak--field approximation, for an inertial observer far
from the source, one can measure the circular orbital speed
as~\cite{Lidsey,saffari}
\begin{equation}\label{v}
v = \frac{{rd\phi }}{{\sqrt {{e^{a\left( r \right)}}} dt}}.
\end{equation}
While using relation (\ref{const}), one can rewrite relation
(\ref{v}) in terms of the conserved quantities as
\begin{equation}\label{v1}
v=\frac{{\sqrt {{e^{a\left( r \right)}}} J}}{{rE}},
\end{equation}
that, by inserting Eq. (\ref{geodesic2}), leads to
\begin{equation}\label{v2}
{v^2} = \frac{{ra'}}{2}.
\end{equation}
In this regard, for the specified minimal coupling model, from
relation (\ref{eax}) in the galactic halo, we have
\begin{equation}\label{aa}
a' = \frac{{\beta  + 2 - \gamma n}}{r}.
\end{equation}
Hence for the model, the tangential velocity reads
\begin{equation}\label{36}
{v^2} = \frac{{\beta  + 2 - \gamma n}}{2},
\end{equation}
where the value of the tangential velocity of test particles in
the circular stable orbits around galactic halo is in the range of
the observed flat rotation curves, i.e. approximately $200-500$
km/s~\cite{1}. Thus, by considering the range of the tangential
velocity $ v $ in a typical spiral galaxy\rlap,\footnote{Note
that, $ v $ is measured in units of $ c $.}\
 it is known that $ v^{2} \approx
\mathcal{O}(10^{-6})$. On the other hand, in order to remain
closed to GR, as mentioned after relation (\ref{assum}), the
$\beta$ parameter must be much smaller than $1$. In fact, the
observed data from the galactic rotation curves indicates that the
$\beta$ parameter is approximately equal to the second power of
tangential velocity\rlap,\footnote{In Ref.~\cite{sobouti}, it has
been shown that the parameter $ \beta $ should depend on the mass
of the gravitating body because any localized matter manifests no
characteristic other than its mass when is sensed from far
distances.}\
 i.e. $\beta \approx v^{2} $~\cite{sobouti}. We have plotted
relation (\ref{36}) in Fig.~$1$ to obtain the allowed values of
the $ n $ parameter in the specified minimal coupling model of
$f(R,T)$ gravity.

\begin{figure}[h]
\begin{center}
\includegraphics[scale=0.8]{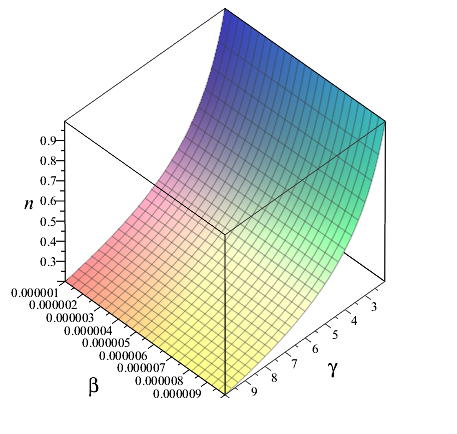}
\caption{The plot of relation (\ref{36}) with $\gamma >2$, $v=300
$ km/s and $ 10^{-6}< \beta < 10^{-5}$.}
\end{center}
\end{figure}

Fig.~$1$ dictates that $ \gamma n\approx 2 $, where in turn,
relation (\ref{roint}) gives the interaction density as
\begin{equation}\label{rho2}
{\rho ^{\left[ {\operatorname{int} } \right]}} \propto {r^{ - 2}}.
\end{equation}
In this case, the metric components (\ref{ebx}) and (\ref{eax})
are
\begin{equation}\label{eb2}
{e^{b\left( r \right)}} = \frac{\beta }{{\alpha n}}={\rm const.}
\end{equation}
and
\begin{equation}\label{ea2}
{e^{a\left( r \right)}} = \frac{{\alpha n}}{{\beta
{{r^{\beta}_{B}} }}}{r^\beta },
\end{equation}
that, as has been set, yield
$e^{a(r)}e^{b(r)}=\left(r/r_B\right)^{\beta}\approx 1 + \beta \ln
\left( {r/r_B} \right)$ for very small $\beta $. On the other
side, by integrating Eq. (\ref{field eq1}), we get
\begin{equation}\label{e(-b)}
{e^{ - b\left( r \right)}} = 1 - \frac{{2G}}{r}\int {4\pi \left( {{\rho ^{\left[ m \right]}} +
{\rho ^{\left[ {{\mathop{\rm int}} } \right]}}} \right)} {{r'}^2}dr'=
 1 - \frac{{2G}}{r}\left( {{M^{\left[ m \right]}} + {M^{\left[ {{\mathop{\rm int}} } \right]}}} \right).
\end{equation}
As it is plausibly assumed that, in the inner parts of the galaxy
(up to $ r={r_{B}}$), the baryonic matter is dominant and that, in
this radius, the mass of the baryonic matter has a fixed value, at
the boundary of the baryonic mass with dark matter, the metric
component $ e^{-b(r)} $ in relation (\ref{e(-b)}) is, therefore,
a constant. Hence, Eq. (\ref{e(-b)}) reads
\begin{equation}\label{s}
{e^{ - b\left( r_{B} \right)}} \approx 1 - \frac{{2G}}{r_{B}}{{\left. {{M^{[m]}}} \right|_{r = r_{B}}}}.
\end{equation}
Furthermore, in the galaxy halo (i.e., in the range $
{r_{B}}<r<r_{D}$, where $ r_{D} $ is the radius wherein the halo
terminates\footnote{That is, the distance, up to which the
tangential velocity remains constant, is considered as the radius
of the dark matter, i.e., up to where the rotation curves are
flat.}\/), the metric component (\ref{e(-b)}) reduces to
\begin{equation}\label{r}
{e^{ - b\left( r \right)}} \approx 1 - \frac{{2G}}{r}{M^{\left[
{{\mathop{\rm int}} } \right]}}.
\end{equation}
Now, due to the continuity of the metric components in the flat
rotation curve boundary of a galaxy, one can set relation
(\ref{s}) equal to relation (\ref{r}) -- while considering the
constancy of relation (\ref{eb2}) -- to achieve
\begin{equation}\label{intr}
{M^{\left[ {\operatorname{int} } \right]}} = \frac{r}{r_{B}}{\left. {{M^{\left[ m \right]}}} \right|_{r = r_{B}}}.
\end{equation}
That is, $ M^{[\rm int]} $ varies linearly with $ r $, which is
consistent with $ \rho^{[\rm int]}\propto r^{-2} $ behavior.
Moreover, we can obtain relation (\ref{e(-b)}) as a constant in
the dark--matter--dominated area.

Another link with the observation is the effect of dark matter on
the light--deflection angle, which is considered in the next
section for the assumed $ f(R,T) $ model.

\section{Light--Deflection Angle}
\indent

This section investigates the light--deflection angle as an effect
of dark matter for the specified minimal coupling model. The
deflection angle $ \Delta\phi  $ is given by
\begin{equation}
\Delta\phi  = 2\left| {\phi \left( {{r_0}} \right) - \phi \left( \infty  \right)} \right| - \pi ,
\end{equation}
where $ r_{0} $ is the radius of the closest approach
to the center of the galaxy. The geodesic equation, Eq. (\ref{geodesic}),
for a photon reduces to~\cite{Weinberg}
\begin{equation}\label{angle}
\phi \left( {{r_0}} \right) - \phi \left( \infty  \right) =
{\int_{{r_0}}^{\infty}  {{e^{\frac{{b \left( r
\right)}}{2}}}\left[ {{e^{a\left( {{r_0}} \right) - a\left( r
\right)}}{{ \left( {\frac{r}{{{r_0}}}} \right)}^2} - 1} \right]}
^{ - \frac{1}{2}}}\frac{{dr}}{r}.
\end{equation}
Using relations (\ref{eb2}) and (\ref{ea2}), we can rewrite
relation (\ref{angle}) as
\begin{equation}\label{46}
\begin{gathered}
  \phi \left( {{r_0}} \right) - \phi \left( \infty  \right) = \sqrt {\frac{\beta }{{\alpha n}}}
  {\int_{{r_0}}^{{r_D}} {\left[ {{{\left( {\frac{{{r_0}}}{r}} \right)}^{\beta -2 }} - 1}
  \right]} ^{ - \frac{1}{2}}}\frac{{dr}}{r} \hfill \\
 \qquad\qquad\qquad\quad+ \int_{{r_D}}^{{\infty }} {\frac{1}{{\sqrt {1 - \frac{{2GM}}{r}} }}}
 {\left[ {\frac{{1 - \frac{{2GM}}{{{r_0}}}}}{{1 - \frac{{2GM}}{r}}}{{\left( {\frac{r}{{{r_0}}}}
 \right)}^2} - 1} \right]^{ - \frac{1}{2}}}\frac{{dr}}{r}, \hfill \\
\end{gathered}
\end{equation}
where, in the first integral term, we have considered $ r_{0} $ in
the region of the flat rotation curves, i.e. $ r_{B}\leqslant
r_{0}<r_{D} $.

One can now exactly integrate the first term in relation
(\ref{46}), which corresponds to the dark--matter--dominated area
(in the region of the flat rotation curve). The second integral
term in relation (\ref{46}) relates to the exterior region of the
dark--matter halo, wherein we integrate it with the Schwarzschild
metric while using the Robertson expansion~\cite{Weinberg}. Hence,
we achieve
\begin{equation}\label{total}
\begin{gathered}
\begin{gathered}
\begin{gathered}
\Delta \phi  = 2\left| {\sqrt {\frac{\beta }{{\alpha n}}} } \right.\left( {\frac{2}{{2 - \beta }}}
\right)\arctan \left( {\sqrt {{{\left( {\frac{{{r_0}}}{{{r_D}}}} \right)}^{\beta  - 2}} - 1} } \right)\\
   \qquad\qquad+ \left. {\arcsin \left( {\frac{{{r_0}}}{{{r_D}}}} \right) + \frac{{GM}}{{{r_0}}}
   \left[ {2 - \sqrt {1 - {{\left( {\frac{{{r_0}}}{{{r_D}}}} \right)}^2}}  - \sqrt {\frac{{{r_D} -
   {r_0}}}{{{r_D} + r_{0}}}} } \right]} \right| - \pi .  \hfill \\
\end{gathered}
\end{gathered}
\end{gathered}
\end{equation}
The variation of the light--deflection angle, as a function of $
r_{0}/r_{D} $ in the galactic halo, has been plotted in Fig.~$2$
(left), with the data from the NGC~$5533$ galaxy with $\beta=1.4
\times10^{-6} $, the NGC~$4138$ galaxy with $
\beta=0.48\times10^{-6} $ and the UGC~$6818$ galaxy with $
\beta=0.12\times10^{-6} $, and all with $ n=1/2 $ and $\alpha=1 $.
For the value of $r_0$ in the second and third terms of relation
(\ref{total}) (which have emerged from the second integral term of
relation (\ref{46})), we have plausibly considered $r_0\approx
r_D$ in the relevant region, see, e.g., Ref.~\cite{sobouti}.
However, when $r_0=r_D$, Eq. (\ref{total}) yields  $\Delta \phi=
4GM/r_D$ as GR, which is consistent with the observation in this
range. Note that, the value $ n=1/2 $ corresponds to a particular
minimal coupling model that satisfies the Bianchi identity for the
Einstein tensor on the cosmological scale, see, e.g.,
Refs.~\cite{shabani13,Zare}. Fig.~$2$ (left) indicates that the
deflection angle can be appreciable as $ {r_0} \ll r_{D} $.
Moreover, we have plotted the light--deflection angle for the
NGC~$4138$ galaxy with $\alpha=1 $, but different values of $ n$,
in Fig.~$2$ (right). The figure shows that by decreasing the value
of $ n $, i.e., by increasing $\gamma $, the deflection angle
increases for each fixed value of $ r_0/r_D $.

 \begin{figure}[h]
\begin{center}
\includegraphics[scale=0.7]{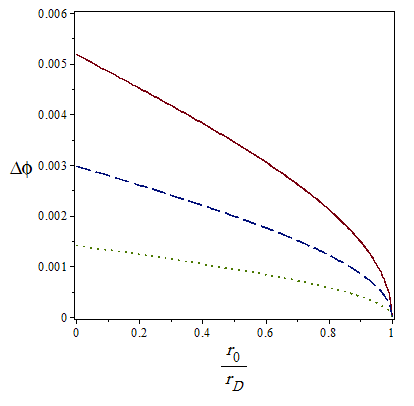}
\includegraphics[scale=0.7]{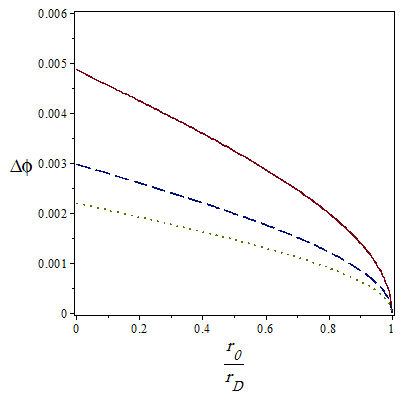}
\caption{ The panels illustrate the light--deflection angle [Eq.
(\ref{total})] for (left) the galaxies NGC~$ 5533 $ (solid line),
NGC~$4138 $ (dashed line) and UGC~$6818 $ (dotted line) with
$n=1/2$ and the $\beta$ parameter as mentioned in the text, and
(right) the NGC~$4138$ galaxy with $ n=1/5 $ (solid line), $ n=1/2
$ (dashed line) and $ n=20/21 $ (dotted line), and all the curves
with $\alpha =1$.}
\end{center}
\end{figure}

Let us now compare the obtained deflection angle from the minimal
coupling model of $ f(R,T) $ gravity with some other dark--matter
models.

In this regard, first consider a commonly used model of the
galactic dark--matter halo with the matter--density
\begin{equation}\label{isodens}
\rho \left( r \right) = {\rho _0}{e^{ - {{\left( r/r_{c}
\right)}^2}}},
\end{equation}
where ${\rho _0}$ is the halo core density and $ r_{c} $ is the
radius of core. This model is a generalization of the
pseudo--isothermal dark--matter model~\cite{42}. In the
weak--field limit, the light--deflection angle, in this model, is
given by~\cite{43}
\begin{equation}\label{49}
\Delta \phi  = \frac{4GM\left( r \right)}{r}= \frac{4G}{r}\int_0^r
4\pi \rho \left( r'\right){r'}^2 dr',
\end{equation}
where $ M(r) $ is the effective mass of the dark matter inside the
radius $ r $. Throughout the galactic halo, inserting the
matter--density (\ref{isodens}) into Eq. (\ref{49}) gives
\begin{equation}\label{50}
\Delta \phi  = \frac{16\pi G\rho _0}{r_0}\int_0^{r_0} {r'}^2 e^{ -
\left( r'/r_c \right)^2}dr'.
\end{equation}
Now, assuming the halo cutoff $ r_{D}$ to be $ r_{D}=10\, r_{c} $,
where $ \rho(r_{D})=0 $, then Eq. (\ref{50}) leads to
\begin{equation}\label{51}
\Delta \phi  = 16\pi G\rho _0\left( \frac{r_D}{10} \right)^2\left[
 - \frac{1}{2}e^{ - \left( r_0/r_D \right)^2} + \frac{\sqrt \pi
}{4}\left( \frac{r_D}{r_0} \right){\rm erf}\left( \frac{r_0}{r_D}
\right) \right].
 \end{equation}
This variation of the light--deflection angle, relation
(\ref{51}), has been plotted in Fig.~$3$, together with the data
of the NGC $5533$ galaxy with $ r_{D}=72$ kpc, the NGC $4138$
galaxy with $ r_{D}=13$ kpc and the UGC $6818$ galaxy with $
r_{D}=6$ kpc, and all with $ \rho_{0}=10^{-14}$ kg/${\rm m}^3$
(these data are from Refs.~\cite{sobouti,43}).
 \begin{figure}[h]
\begin{center}
\includegraphics[scale=0.7]{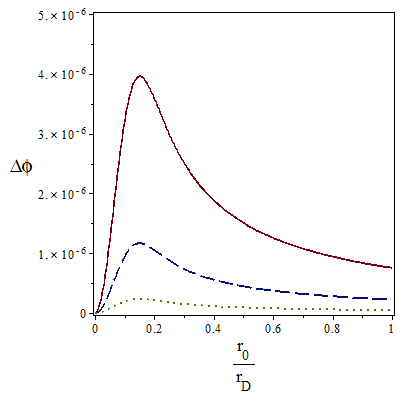}
\caption{The curves illustrate the light--deflection angle [Eq.
(\ref{51})] for galaxies NGC $5533$ (solid line), NGC $4138$
(dashed line) and UGC $6818$ (dotted line) with the parameters
mentioned in the text.}
\end{center}
\end{figure}
Fig.~$3$ illustrates that the deflection angle, in this model,
increases up to $ r_c $ and then, starts to decline in the
dark--matter halo.

The comparison of the deflection angle of our model with the
brane--$ f(R) $ gravity model (proposed in Ref.~\cite{Sefidgar})
indicates that both models have the same behavior; i.e., the
deflection angle reduces in the dark--matter halo with the radius.

\section{Conclusions}
\indent

In this work, we have considered $ f(R,T)$--modified theory of
gravity to explain the dark--matter effects in spiral galaxies as
inferred from the flat rotation curves. The additional term in the
field equations, that is acquired in this type of theory, can be
interpreted as an alternative for the dark--matter effects in the
galaxy. We have obtained the metric components for a spherically
symmetric and static spacetime in the vicinity of general
relativity solutions. However, we have concentrated on a
particular minimal coupling model in this theory, and have derived
the metric components in the galactic halo. Then, we have fixed
the components by the rotational velocities of galaxies for the
model. Finally, we have obtained that the mass corresponding to
the interaction term (which appears in the Einstein modified field
equation) varies linearly with the radius, hence, the interaction
mass can cause a flat rotation curve in the halo of galaxies.

In addition, we have obtained the light--deflection angle, using
the metric components derived for the model, in the galactic halo.
The diagrams for the light--deflection angle consistently indicate
that a galaxy with larger mass has a bigger deflection angle. For
a galaxy with a specified mass, we have plotted the
light--deflection angle for different values of $ n$ (a power that
shows the strength of the effect of the trace of the
energy--momentum tensor of the matter in the specified minimal
coupling model). The figures indicate that by decreasing the value
of $ n $, the light--deflection angle increases. This can be
explained by noting that the density of the baryonic matter, for
smaller values of $ n$, rapidly decreases in the halo. Indeed,
this is due to the fact that the density of the baryonic matter,
for smaller values of $ n $, is more concentrated in the region $
{r_0} \ll r_{D} $ compared to the density of the interaction term.

\section*{Acknowledgments}
\indent

We thank the Research Council of Shahid Beheshti University for
financial support.

\end{document}